\documentstyle[amssymb,preprint,aps,tighten]{revtex}

\begin{document}
\draft
\title{Weakly screened thermonuclear reactions in astrophysical plasmas: Improving
Salpeter's model}
\author{Theodore E. Liolios \thanks{%
http://www.liolios.net}}
\address{Hellenic Naval Academy of Hydra\\
School of Deck Officers, Department of Science\\
Hydra Island 18040, Greece}
\date{January 2003}
\maketitle

\begin{abstract}
This paper presents a detailed study of the electron degeneracy and
nonlinear screening effects which play a crucial role in the validity of
Salpeter's weak-screening model. The limitations of that model are
investigated and an improved one is proposed which can take into account
nonlinear screening effects. Its application to the solar pp reaction
derives an accurate screening enhancement factor and provides a very
reliable estimation of the associated neutrino flux uncertanties.
\end{abstract}


\oddsidemargin -0.25cm \evensidemargin -0.25cm \topmargin -1.0cm \textwidth %
16.3cm \textheight 22.3cm

\section{Introduction}

Thermonuclear reactions constitute the source of stellar energy and all the
laws that govern such reactions deserve to be thoroughly investigated. The
enhancing influence of stellar plasmas on thermonuclear reaction rates has
been studied by many authors (\cite{salpeter}-\cite{lioliosprc2001}, and
references therein) who derive corrective factors (known as Screening
Enhancement Factors:SEFs) by which the reaction rates are multiplied in
order to take into account screening effects. There is a variety of models,
each of which has inherent limitations, while some of them have been the
subject of intense controversy\cite{bahcall5}.

The most widely used screening models are Salpeter's weak screening (WS)
model (S)\cite{salpeter} and Mitler's one (M)\cite{mitler}, the latter
covering the effect at all densities. Actually these models are used in
solar evolution codes giving quantitative estimates of the neutrino flux
uncertainties associated with the screening effect\cite{dzitko,ricci}.
However, there are still sources of uncertainty originating from these
models which have not been dealt with. In a very important study\cite
{adelberger} of solar fusion cross section it was admitted that a
satisfactory analytical investigation of electron degeneracy and nonlinear
screening effects for Salpeter's model is not yet available. This paper
satisfies that need and investigates the relevant effects on solar neutrino
fluxes.

The double aim of this paper is a) to present a detailed study of how
degeneracy affects Salpeter's thermonuclear screening formalism, b) to
quantitatively investigate the nonlinear screening effects disregarded by
Salpeter's model, avoiding at the same time Mitler's arbitrary assumption
that in all stellar plasmas the charge density around the reacting nuclei is
equal to the average electron density in the plasma.

The layout of the paper is as follows:

In Sec. II. there is a very brief introduction to the fundamental
assumptions of the Thomas-Fermi (TF) theory in screened stellar plasmas,
which will be most useful in highlighting the limitations of the existing
models and building a new improved one. In Sec. III. , we derive Salpeter's
SEF for a WS non-degenerate environment from the first principles of the TF
theory, thus avoiding Salpeter's statistical approach (Boltzmann's formula)
which disregards the nonlinear terms of Poisson's equation. The TF theory is
also employed in Sec. IV. where we derive Salpeter's SEF for a partially
degenerate (PD)\ WS environment extracting a new analytical degeneracy
factor which avoids the complicated improper integrals involved in the
formulas currently used in stellar evolution codes and allows a detailed
study of degeneracy effects. In Sec. V. we investigate the electron
degeneracy effects on SEFs defining quantitatively the following
environments: non-degenerate (ND), weakly degenerate (WD), intermediately
degenerate (ID), strongly degenerate (SD) and the completely degenerate (CD)
one. In Sec.VI. , we prove that the definition of the WS limit (used by both
the S and M models), which actually yields the Debye-Huckel (DH) potential,
forbids the use of that potential inside the tunneling region and the
subsequent derivation of the S model. The region of validity of the DH
potential in stellar plasmas is obtained by defining a tuning equation which
can tune the accuracy of the weak-screening limit at will. In Sec. VII. we
design a new model for weakly screened thermonuclear reactions which is
capable of taking into account the error associated with nonlinear screening
effects. In Sec. VIII. we apply the new model to the solar pp reaction and
estimate the associated neutrino flux uncertainties. Finally, in Sec. IX.,
there is a thorough discussion of the new models and their results

\section{The Thomas-Fermi model in screened astrophysical plasmas}

Let us assume that in a completely ionized stellar plasma each atom of
element $_{Z}^{A}M$ releases $\left( Z_{i}+1\right) $ free particles in the
gas, that is $Z_{i}$ electrons plus the nucleus. The conditions of complete
ionization impose the first constraint on the calculations that follow. We
adopt Salpeter's quantitative interpretation\cite{salpeter} of that
assumption, namely:\thinspace $T_{6}>>0.16Z_{i}^{2}$ for all ions $\left(
i\right) $ in the plasma, where $T_{6}$ is the plasma temperature in million
Kelvins. On the other hand we assume that for a particular thermonuclear
reaction ignition does occur and since hydrogen burning starts first we will
consider temperatures higher than the hydrogen ignition one, which, for
unscreened reactions, is roughly \cite{bowers} $T_{6}\gtrsim 10$.

$\,$Let $N_{i}\,$be the number density of element $\left( i\right) $, which
is given as a function of plasma mass density $\rho $, fraction by weight $%
X_{i},$ and mass number $A_{i\,}\,$by the formula $N_{i}=\rho \left(
X_{i}/A_{i}\right) N_{0}$, $N_{0}$ being Avogadro's number. Then the total
number density of free particles will be $N=\sum_{i}N_{i}\left(
Z_{i}+1\right) $ or else $N=\rho N_{0}\sum_{i}\left( X_{i}/A_{i}\right)
\left( Z_{i}+1\right) .$ The number density of electrons $\left( i=e\right) $
can then be written $N_{e}=\rho N_{0}\sum_{i}Z_{i}\left( X_{i}/A_{i}\right) $
$\,$as well as $N_{e}=\rho N_{0}/\mu _{e}$ where we have used the mean
molecular weight per electron $\mu _{e}=2\left( 1+X_{H}\right) ^{-1}$, which
should not be confused with the electronic chemical potential. In a
completely ionized plasma the global charge density $\rho
_{q}=\sum_{i}N_{i}\left( Z_{i}e\right) $ \thinspace will be

\begin{equation}
\rho _{q}=\sum_{i\neq e}N_{i}Z_{i}e-N_{e}e=0
\end{equation}
which is the quantitative definition of plasma neutrality.

All ions obey the Fermi-Dirac statistics and therefore the distribution of
ionic momenta in thermal equilibrium will be:

\begin{equation}
n_{i}\left( p\right) =\frac{8\pi p^{2}/h^{3}}{\exp \left[ \left( \frac{p^{2}%
}{2m_{i}}-\mu _{i}\right) /kT\right] +1},\qquad \int_{0}^{\infty
}n_{i}\left( p\right) dp=N_{i}  \label{fdstat}
\end{equation}
where $n_{i}\left( p\right) dp\,$is the number of ions $\left( i\right) \,$%
with momenta lying within $\left[ p,p+dp\right] $ and $\mu _{i}$ is the
corresponding chemical potential defined by Eq. $\left( \ref{fdstat}\right) $%
.

Plasma neutrality then reads:

\begin{equation}
\sum_{i\neq e}Z_{i}\int_{0}^{\infty }n_{i}\left( p\right)
dp-\int_{0}^{\infty }n_{e}\left( p\right) dp=0  \label{plneutr}
\end{equation}
where $n_{e}$ is the electron number density of all electrons at distance $r$
from the nucleus with total energy $\varepsilon $ and momentum $p\left(
r\right) $ , respectively.

After the (virtual) introduction of the point nucleus $Z_{0}e\,$in the
plasma the charge neutrality condition no longer holds in the polarized
vicinity of the nucleus and now we need to define a local charge density
given by

\begin{equation}
\widetilde{\rho }_{q}\left( r\right) =\sum_{i}\widetilde{N}_{i}\left(
Z_{i}e\right) =\int_{0}^{\infty }\left[ \sum_{i\neq e}\widetilde{n}%
_{i}\left( p\right) Z_{i}e-\widetilde{n}_{e}\left( p\right) e\right] dp
\end{equation}
which is actually a function of distance from the charge $Z_{0}e$. The
perturbed quantities from now on will be denoted by tilded letters. The
local number densities (denoted now by $\widetilde{n}_{e}$ and $\widetilde{n}%
_{i}$) will be modified in the sense that their chemical potential will be
shifted by $-eZ_{i}\Phi $ for nuclei and $e\Phi $ for electrons. Note that
plasma charge neutrality still holds so that the total charge $Q_{tot}\,$%
defined by

\begin{equation}
Q_{tot}=\int_{0}^{\infty }\widetilde{\rho }\left( r\right) 4\pi
r^{2}dr+Z_{0}e
\end{equation}
is of course $Q_{tot}=0.\,$

The self-consistent screened Coulomb potential acting in the vicinity of the
polarizing nucleus is given by Poisson's equation

\begin{equation}
\nabla ^{2}\Phi \left( r\right) =-4\pi e\int_{0}^{\infty }\left[ \sum_{i\neq
e}\widetilde{n}_{i}\left( p\right) Z_{i}e-\widetilde{n}_{e}\left( p\right)
e\right] dp  \label{poisson}
\end{equation}
The boundary conditions for the above equation are

\begin{equation}
\lim_{r\rightarrow 0}\Phi \left( r\right) =\frac{Z_{0}e}{r}%
,\lim_{r\rightarrow \infty }\Phi \left( r\right) =0  \label{bcond}
\end{equation}
while the perturbed electron and ion densities will now be given by

\begin{equation}
\widetilde{n}_{e}\left( p\right) =\frac{8\pi p^{2}/h^{3}}{\exp \left\{
\left[ \frac{p^{2}}{2m_{e}}-\left( \mu _{e}+e\Phi \right) \right]
/kT\right\} +1}
\end{equation}
and

\begin{equation}
\widetilde{n}_{i}\left( p\right) =\frac{8\pi p^{2}/h^{3}}{\exp \left\{
\left[ \frac{p^{2}}{2m_{i}}-\left( \mu _{i}-eZ_{i}\Phi \right) \right]
/kT\right\} +1}
\end{equation}
Our study will be greatly facilitated by the introduction of the degeneracy
parameter $a_{i}$ for each ion defined as $a_{i}=-\mu _{i}/kT$ , and $%
\widetilde{a}_{i}=-\left( \mu _{i}-Z_{i}e\Phi \right) /kT=a_{i}+eZ_{i}\Phi
/kT,$ respectively. Note that for electrons $\left( i=e\right) $ we have
assumed that $Z_{e}=-1.$ The degeneracy parameter\cite{claytonbook} gives us
the degree of degeneracy of a fermionic gas in a very transparent way. Using
the Fermi-Dirac function of one-half which is defined as

\begin{equation}
F_{1/2}\left( a\right) =\int_{0}^{\infty }\frac{u^{1/2}du}{\exp \left(
a+u\right) +1}
\end{equation}
the number densities can now be written

\begin{equation}
N_{i}\left( a\right) =\frac{4\pi }{h^{3}}\left( 2m_{i}kT\right)
^{3/2}F_{1/2}\left( a\right)  \label{nuden}
\end{equation}

\section{Non-degenerate, weakly screened, non-relativistic environment.}

For $a_{i}>0$ the electron gas is either in a WD or in ND state. Then we can
write\cite{claytonbook}

\begin{equation}
F_{1/2}\left( a_{i}\right) =-\frac{\sqrt{\pi }}{2}\sum_{n=1}^{\infty }\frac{%
\left( -1\right) ^{n}e^{-na_{i}}}{n^{3/2}}  \label{f12ws}
\end{equation}
Let us further assume that there are some critical values $%
a_{e}^{*},a_{i}^{*}$ so that for all degeneracy parameters which satisfy the
conditions

\begin{equation}
a_{e}>a_{e}^{*}>0,\,\,a_{i}>a_{i}^{*}>0  \label{conda}
\end{equation}
the second term of the series $\left( n=2\right) $ in Eq. $\left( \ref{f12ws}%
\right) $ is negligible. Then obviously for all these parameters we can
retain only the first term of the above expansion for $F_{1/2},$ thus
introducing the first linearization of the TF model which yields the
following number densities:

\begin{equation}
N_{e}\left( r\right) \simeq N_{e}\left( a_{e}\right) \exp \left( \frac{e\Phi
\left( r\right) }{kT}\right) ,\,\,\,N_{i}\left( r\right) \simeq N_{i}\left(
a_{i}\right) \exp \left( \frac{-Z_{i}e\Phi \left( r\right) }{kT}\right)
\label{boltzmann}
\end{equation}
Under such conditions ions are always non-degenerate since $a_{e}\ll a_{i},$
so we should focus solely on electron degeneracy from now dropping the index 
$e$ when referring to the electron degeneracy parameter. The error committed
in the calculation of the electron density by retaining only the first term
in the expansion of $F_{1/2}\left( a\right) $ depends on the first neglected
term, that is $\exp \left( -2a^{*}\right) /2^{3/2}\,,$\thinspace thus for $%
a^{*}=2$ the error is less than $1\%$. As will soon become obvious this ND
limit is not enough for the WD central solar plasma where the degeneracy
parameter is smaller. Note that Eqs. $\left( \ref{boltzmann}\right) $ are
actually Boltzmann's formulas so that we can now clearly state that
Boltzmann's statistics is valid for the description of a fermionic gas when $%
a_{i}\geq 2.$

The second linearization is actually based on the WS limit which is defined
as

\begin{equation}
\frac{e\Phi \left( r\right) }{kT}\ll 1,\,\,\frac{Z_{i}e\Phi \left( r\right) 
}{kT}\ll 1  \label{condwes}
\end{equation}
Note that these conditions, which are supposed to be valid at all relative
distances between ions will soon be proved to be abused when used in the
derivation of the M and S SEFs.

Therefore, in a (WS, ND) stellar environment, disregarding nonlinear terms
in Eqs. $\left( \ref{boltzmann}\right) ,\,$ we have

\begin{equation}
N_{e}\left( r\right) \simeq N_{e}\left( a\right) \left( 1+\frac{e\Phi \left(
r\right) }{kT}\right) ,N_{i}\left( r\right) \simeq N_{i}\left( a_{i}\right)
\left( 1-\frac{Z_{i}e\Phi \left( r\right) }{kT}\right)  \label{numdens}
\end{equation}
and of course Eq. $\left( \ref{poisson}\right) $ will be

\begin{equation}
\nabla^{2}\Phi \left( r\right) =-4\pi e\left[ \sum_{i\neq e}N_{i}\left(
a_{i}\right) \left( 1-\frac{Z_{i}e\Phi \left( r\right) }{kT}\right)
Z_{i}-N_{e}\left( a\right) \left( 1+\frac{e\Phi \left( r\right) }{kT}\right)
\right]
\end{equation}
Using the plasma neutrality condition we obtain

\begin{equation}
\nabla^{2}\Phi \left( r\right) =\frac{4\pi e^{2}}{kT}\left( \sum_{i\neq
e}N_{i}\left( a_{i}\right) Z_{i}^{2}+N_{e}\left( a\right) \right) \Phi
\left( r\right)
\end{equation}
or else

\begin{equation}
\nabla^{2}\Phi \left( r\right) =\frac{\Phi \left( r\right) }{R_{D}^{2}}
\label{fd}
\end{equation}
where

\begin{equation}
R_{D}^{-2}=\frac{4\pi e^{2}}{kT}\left( \sum_{i\neq e}N_{i}\left(
a_{i}\right) Z_{i}^{2}+N_{e}\left( a\right) \right)
\end{equation}
is the usual DH radius.

In order to solve Eq. $\left( \ref{fd}\right) $, we apply the boundary
conditions given by Eq. $\left( \ref{bcond}\right) $ deriving the well-known
DH screened Coulomb potential

\begin{equation}
\Phi _{DH}\left( r\right) =\frac{Z_{0}e}{r}\exp \left( -r/R_{D}\right)
\label{potdh}
\end{equation}
In such a case the SEF for the thermonuclear reaction $%
_{Z_{1}}^{A_{1}}M+_{Z_{0}}^{A_{0}}M\,,$ is given as usual by Salpeter's
formula\cite{salpeter}

\begin{equation}
f_{s}=\exp \left( \frac{Z_{1}Z_{0}e^{2}}{kTR_{D}}\right)  \label{fs}
\end{equation}

Note that Salpeter's SEF was derived on two consecutive linearization steps
which are both based on the WS condition given by Ineq. $\left( \ref{condwes}%
\right) .$ The first step is taken in the actual derivation of the DH
potential via Eqs. $\left( \ref{numdens}\right) $ while the second one is
taken when nonlinear terms of the DH potential expansion are disregarded
inside the integral of Eq. $\left( \ref{wkbint}\right) .$ The second step
has been exhaustively studied by the author$\cite{lioliosprc2000}$ while the
first one, whose validity is crucial to the validity of the second one, will
be investigated in the present paper.

\section{Partially degenerate, weakly screened, non-relativistic environment.
}

Let us now assume a partially degenerate (PD) environment where $a<0$ . In
such a case the Fermi-Dirac functions can be simplified as follows\cite
{claytonbook}:

\begin{equation}
F_{1/2}\left( a\right) =\frac{2}{3}\left( -a\right) ^{3/2}\left( 1+\frac{\pi
^{2}}{8a^{2}}+\frac{7\pi ^{4}}{640a^{4}}...\right)  \label{f12deg}
\end{equation}
which are very accurate when $a<-3$ as they only commit an error of less
than 1\% in the calculation of electron density.

Again we would only be interested in electron degeneracy as nuclei are
almost always non-degenerate. Therefore for nuclei we still have

\begin{equation}
N_{i}\left( r\right) \simeq N_{i}\left( a_{i}\right) \exp \left( \frac{%
-Z_{i}e\Phi \left( r\right) }{kT}\right)
\end{equation}
As for electrons, setting $y=-\left( e\Phi /akT\right) >0$ we can write

\begin{equation}
F_{1/2}\left( \widetilde{a}\right) \simeq \frac{2}{3}\left[ -\left(
a+y\right) \right] ^{3/2}\left[ 1+\frac{\pi ^{2}}{8}\left( a+y\right) ^{-2}+%
\frac{7\pi ^{4}}{640}\left( a+y\right) ^{-4}\right]  \label{f12a}
\end{equation}
where we have assumed that $0<y\ll 1$ so that $y^{2}\simeq 0.$ Retaining
only terms up to $a^{-4}$ in the truncated form of $F_{1/2}\left( a\right) $
, after some tedious algebra the previous equation yields 
\begin{equation}
F_{1/2}\left( \widetilde{a}\right) \simeq F_{1/2}\left( a\right) \left[
1+\theta \left( a\right) \frac{e\Phi }{kT}\right]
\end{equation}
where the degeneracy parameter $0<\theta \left( a\right) <1$ is given by

\begin{equation}
\theta \left( a\right) =\frac{3}{2}\left( -a\right) ^{-1}-\frac{2}{3}\frac{%
\left( -a\right) ^{1/2}}{F_{1/2}\left( a\right) }\left( \frac{\pi ^{2}}{%
4a^{2}}+\frac{7\pi ^{4}}{160a^{4}}\right)  \label{dp0}
\end{equation}

Being consistent with our initial approximation of $F_{1/2}\left( a\right) ,$
when $a<-3$ we derive a very simple and handy formula, namely:

\begin{equation}
\theta \left( a\right) =-\frac{5}{2a}\frac{384a^{4}-16\pi ^{2}a^{2}-7\pi ^{4}%
}{640a^{4}+80\pi ^{2}a^{2}+7\pi ^{4}}  \label{dp1}
\end{equation}
If alternatively we keep only terms up to $a^{-2}$ (just as in Ref.$\cite
{shiv}$) then we obtain

\begin{equation}
\theta \left( a\right) =\frac{-\left( 24a^{2}-\pi ^{2}\right) }{3a\left(
8a^{2}+\pi ^{2}\right) }  \label{dp2}
\end{equation}
The electron number density now reads

\begin{equation}
N_{e}\left( r\right) =N_{e}\left( a\right) \left[ 1+\theta \left( a\right) 
\frac{e\Phi }{kT}\right]
\end{equation}
which replaces its ND form given by Eq. \ref{numdens}. Applying the same
procedure as in the ND case the DH radius in PD conditions assumes the usual
form

\begin{equation}
R_{D}^{-2}=\frac{4\pi e^{2}}{kT}\left( \sum_{i\neq e}N_{i}\left(
a_{i}\right) Z_{i}^{2}+\theta \left( a\right) N_{e}\left( a\right) \right)
\end{equation}
where now degeneracy is taken into account through Eq. $\left( \ref{dp0}%
\right) $ or its much simpler form Eq.$\left( \ref{dp1}\right) $ in the
right degeneracy conditions$.$

The quantity $\theta \left( a\right) $ is the degeneracy factor $\left(
F_{1/2}^{^{\prime }}/F_{1/2}\right) $ calculated numerically in Ref.\cite
{salpeter} given here in an approximate analytic form which allows: A) a
transparent study of electron degeneracy effects in stellar plasmas and , B)
a very fast and easy way of calculating degeneracy effects in PD stellar
gases. Note that the calculation of the quantity $\left( F_{1/2}^{^{\prime
}}/F_{1/2}\right) $ delays the computing process in stellar evolution codes
due to the improper integrals involved. On the other hand Mitler had to
''guess''\thinspace his form of the degeneracy factor while here it is
derived in a rigid mathematical method.

Our novel degeneracy parameter has the right limit for non-relativistic CD
gases, just as noted in Ref.\cite{salpeter}, namely

\begin{equation}
\lim_{a\rightarrow -\infty }\theta \left( a\right) =\frac{3}{2}\left(
-a\right) ^{-1}=5T_{6}\left( \frac{\rho }{\mu _{e}}\right) ^{-2/3}
\label{dpfermi}
\end{equation}
Admittedly, Mitler's ingenious guess\cite{mitler}

\begin{equation}
\theta \left( a\right) =\left[ 1+\frac{4}{9}\left[ \frac{3}{2}F_{1/2}\left(
a\right) \right] ^{4/3}\right] ^{-1/2}  \label{dpmit}
\end{equation}
is correct within 2\% over the entire stellar profile, just as he remarks.
However, the main shortcomings of his formula are A) It has been guessed and
lacks mathematical rigor contrary to the present one, which is based on a
well established mathematical mechanism, B) It taxes heavily the computing
process especially in SD gases while ours, in the same regime, is a rational
algebraic function.

In Fig. 1, we plot all available analytic formulas for the degeneracy
parameter $\theta \left( a\right) ,$ namely Eqs. $\left( \ref{dp0}\right)
,\left( \ref{dp1}\right) ,$ $\left( \ref{dp2}\right) $, $\left( \ref{dpfermi}%
\right) ,$ as well as Mitler's formula given by Eq.$\left( \ref{dpmit}%
\right) $. Mitler's guessed formula will serve as an index of accuracy for
our formulas. As regards the new formulas for degeneracy, in the region $%
-2<a $, we observe a rapid decrease of $\theta \left( a\right) $ although we
know that for ND environments $\theta \left( a\right) $ is supposed to tend
to unity. This region is naturally outside the limits of our formulas and is
signified by the first vertical bar in Fig.1. On the other hand it is
obvious that for values $a<-10$ electron degeneracy is practically complete
and then the simple Eq.$\left( \ref{dpfermi}\right) $ should be used. The
boundary of that completely degenerate regime (CD) is represented by the
third vertical bar (the regions corresponding to the intermediate (ID) and
strong degeneracy (SD) regime that are shown in Fig. 1 will be defined in
the next section). Note that as we remarked at the beginning of this section
the truncated form of $F_{1/2}\left( a\right) $ \thinspace yields an
accurate number density provided $a<-3.$ This is now corroborated by the
degeneracy factor of our model which are shown in Fig. 1. and is indeed very
accurate for $a<-3.$ In the analysis of Fig. 1 special attention should be
paid to the dash-dot-dotted curve corresponding to Eq. $\left( \ref{dp2}%
\right) $ as it is closely related to an important relevant paper$\cite{shiv}
$. That paper derives screened Coulomb potentials for single nuclei placed
in a sea of electrons. If we neglect the presence of all other ions but the
central one, and keep only $a^{-2}$ terms in our study then we recover the
screening radius obtained in Ref.$\cite{shiv}.$ This coincidence enables us
to make full use of the results of that work since we can study
thermonuclear reactions in stellar plasmas by adding to their purely
electronic screening radii the effects of a ND non-relativistic sea of
positive (spectator) ions. However, as one can observe from Fig.1, if we
follow the assumption of Ref.$\cite{shiv}$ retaining only terms of order $%
a^{-2}$ in the truncation of Eq. $\left( \ref{f12deg}\right) $ then we
derive a very poor description of the degeneracy effects (i.e. Eq. $\ref{dp2}
$). Therefore the study of thermal and relativistic effects in dielectric
screening given in Ref.$\cite{shiv}$ is not valid for weakly ND regimes as
the authors claim but for extremely degenerate regimes, where of course
their thermal correction $\left( 1-\pi ^{2}/24a^{2}\right) $ to the cold
electron gas density is negligible. The present paper improves their study
and produces really significant thermal corrections. Namely, the quantity $%
\left( 1-\pi ^{2}/24a^{2}\right) $ appearing in Ref. \cite{shiv} should be
replaced by the much more accurate one derived in this paper:

\begin{equation}
\frac{3}{2}\left( -a\right) ^{-3/2}F_{1/2}\left( a\right) -\frac{2}{3}\left( 
\frac{\pi ^{2}}{4a^{2}}+\frac{7\pi ^{2}}{160^{4}}\right)  \label{newshiv}
\end{equation}
It is straightforward to prove that Eq. $\left( \ref{newshiv}\right) $
coincides with $\left( 1-\pi ^{2}/24a^{2}\right) $ if we only keep terms of
order $a^{-2}.$

We will now attempt to establish quantitatively the true limits of
Salpeter's and Mitler's formulas in terms of degeneracy and non-linear
screening effects

\section{Degeneracy effects}

Since the value of $a$ is related to the stellar environment through the
following relation\cite{claytonbook}

\begin{equation}
\log _{10}\left( \frac{\rho }{\mu _{e}}T^{-3/2}\right) =\log
_{10}F_{1/2}\left( a\right) -8.044  \label{adt}
\end{equation}
we can write for all values $\left( a\geq a^{*}\right) $ the following
inequality:

\begin{equation}
\frac{\rho }{\mu _{e}}\leq 10^{9+Q\left( a^{*}\right) }T_{6}^{3/2}
\label{ineqa}
\end{equation}
where $Q\left( a^{*}\right) =\log _{10}F_{1/2}\left( a^{*}\right) -8.044.$

Obviously using Ineq. $\left( \ref{ineqa}\right) $ we can define any
degeneracy limit in an $\left( \rho ,T\right) $ plane. Therefore, according
to the previous discussion, it is plausible to define the limit between ND
and WD environments at $a^{*}=2.\,$

Inserting the value $a^{*}=2$ into Ineq. $\left( \ref{ineqa}\right) $ we can
derive here in a transparent way the boundaries of the WD and the ND\
(weakly screened) environments

\begin{equation}
\frac{\rho }{\mu _{e}}\leq T_{6}^{3/2}  \label{ineqnd}
\end{equation}
In ND gases $\left( a>2\right) $ no correction is needed for degeneracy
while in WD ones $\left( -3<a<2\right) \,$the electron density should
inevitably be multiplied by the usual value of $\theta \left( a\right) \,$%
given by\cite{salpeter}

\begin{equation}
\theta \left( a\right) =F_{1/2}^{-1}\left( a\right) \frac{dF_{1/2}\left(
a\right) }{da}  \label{dpf12derf12}
\end{equation}
or Mitler's Eq. $\left( \ref{dpmit}\right) ,\,$alternatively.

For instance for central solar condition $\left( \rho /\mu _{e}\simeq
148,T_{6}\simeq 15.6\right) $ the degeneracy parameter is $a=1.1$ while in
the solar region of maximum energy production (i.e. at a distance of $0.075$
solar radii where $T_{6}=14$ and $\rho =105\,g/cm^{3}),$ we have $a=1.3.$
Obviously, the sun lies in the WD zone of the equation of state.

On the other hand, as we have shown, the strong degeneracy (SD)\ limit will
be obeyed if $a\leq -5$ that is

\begin{equation}
\frac{\rho }{\mu _{e}}\geq 71T_{6}^{3/2}  \label{ineqsd}
\end{equation}
and wherever this condition is valid we can use Eq. $\left( \ref{dp1}\right) 
$ to take into account degeneracy effects.

As is obvious from Fig.1, for $a\leq -10$ the electron gas is completely
degenerate and of course under such conditions

\begin{equation}
\frac{\rho }{\mu _{e}}\geq 193T_{6}^{3/2}
\end{equation}
In that region Eq. $\left( \ref{dp1}\right) \,$can be replaced by the much
simpler Eq. $\left( \ref{dpfermi}\right) $

The intermediate degeneracy (ID) limit is a straightforward result of the
condition that the electron pressure in a ND gas is equal to the electron
pressure in a CD gas, namely

\begin{equation}
\frac{\rho }{\mu _{e}}=24T_{6}^{3/2}  \label{ineqdeg}
\end{equation}
In such a case the degeneracy parameter which defines this limit is $a=-2$

The limit for a fully relativistic degenerate electron gas\cite{claytonbook}
will define the boundaries of the non-relativistic nature of our study

\begin{equation}
\frac{\rho }{\mu _{e}}>7.3\times 10^{6}  \label{relativistic}
\end{equation}
Let us now assume a zero-metallicity hydrogen-helium plasma where we need to
examine the effects of degeneracy on Salpeter's formula when applied to pp
reactions. Thus, we assume that the target nucleus whose screened Coulomb
potential we need to derive is a proton (i.e. $Z_{0}=1)$. Then, the usual
definition of the weak screening condition (see Ineq. $\left( \ref{condwes}%
\right) $)$\,$which actually yields the S model is written

\begin{equation}
\frac{Z_{i}Z_{0}e^{2}}{\left\langle r\right\rangle _{i}}\exp \left( -\frac{%
\left\langle r\right\rangle _{i}}{R_{D}}\right) \ll kT  \label{misi}
\end{equation}
where $i=H,He$ and $\left\langle r\right\rangle _{i}$ the interionic
distance between $H-H,H-He$,\thinspace respectively. The average distance is
easily shown to be of the same order of magnitude as the inter-electronic
one. Since $\left\langle r\right\rangle _{ee}\sim n_{e}^{-1/3},$and $%
n_{e}=\rho N_{0}/\mu _{e}\,$we can use Eq. $\left( \ref{misi}\right) $ to
derive the following limits:

For $i=H$ we have

\begin{equation}
\frac{\rho }{\mu _{e}}\ll 350T_{6}^{3}
\end{equation}
while for $i=He$

\begin{equation}
\frac{\rho }{\mu _{e}}\ll \frac{350}{8}T_{6}^{3}
\end{equation}
In Fig. 2, we map all the degeneracy zones of the equation of state of an
electron gas. Various stellar environments are shown in the map such as that
of main sequence stars (e.g. the sun), neutron stars, and the ones
encountered during explosive hydrogen burning and x-ray bursts. In the top
left-hand-side corner we observe two lines which represent the boundaries of
validity of the usual WS assumption given by Eq. $\left( \ref{misi}\right) $
, which leads to the derivation of the S model. If we drop the
zero-metallicity scenario, new lines will appear parallel to the $%
H-H,H-He,\, $such as $H-C,H-N,$ and so on. The larger the charge of the
metal included the further its line will be to the right of the two ones
shown in the map. Likewise, if we study other reaction with $Z_{0}>1$ we
will observe an even more pronounced shift to the relevant boundaries. To
the far right of all these lines the WS limit assumed by the S model is
supposed to be valid for the description of the screening effects regardless
of how degenerate the electronic environment is. However, even screening
condition $\left( \ref{misi}\right) \,$is abused in very degenerate stars
with a rich composition of heavy metals. It has now become very clear that
the S model cannot be safely used for reactions other than the pp one in PD
main-sequence stars (as often remarked by other authors as well \cite{dzitko}%
).

Since the interelectronic distance is $\left\langle r\right\rangle _{ee}\sim
n_{e}^{-1/3}$ we can easily see that 
\begin{equation}
\left\langle r\right\rangle _{ee}\sim \left[ 1+\theta \left( a\right) \frac{%
e\Phi }{kT}\right] ^{-1/3}
\end{equation}
therefore the phenomenon of degeneracy increases the average distance
between electrons while it obviously reduces the screening effect and thus
thermonuclear energy is released more slowly under degenerate conditions
than under non-degenerate ones. As the energy production rate is directly
proportional to SEFs it would be very interesting to study how that rate
responds to turning off and on the degeneracy effect. Salpeter's SEF can be
written\cite{salpeter}:

\begin{equation}
f_{s}=\exp \left( 0.188Z_{1}Z_{2}\zeta \rho ^{1/2}T_{6}^{-3/2}\right)
\end{equation}
where

\begin{equation}
\zeta =\sqrt{\sum_{i\neq e}\frac{X_{i}Z_{i}^{2}}{A_{i}}+\theta \left(
a\right) \sum_{i\neq e}\frac{X_{i}Z_{i}}{A_{i}}}
\end{equation}
In the most representative case of a zero-metallicity stellar plasma we have

\begin{equation}
\zeta =\sqrt{1+\frac{\theta \left( a\right) }{\mu _{e}}}
\end{equation}
therefore we can write Salpeter's SEF as follows:

\begin{equation}
f_{s}\left( \rho ,T_{6},\mu _{e}\right) =\left[ \exp \left(
0.188Z_{1}Z_{2}\rho ^{1/2}T_{6}^{-3/2}\right) \right] ^{\sqrt{1+\frac{\theta
\left( \rho ,T_{6}\right) }{\mu _{e}}}}  \label{salsef}
\end{equation}
which vividly depicts the influence of electron degeneracy. Note, for
example, that if there was no such think as degeneracy then $\theta \left(
\rho ,T_{6}\right) =1$ and the previous formula coincides with the one
describing ND gases. Therefore, we can plot Salpeter's SEF for various
environments turning on and off electron degeneracy to illustrate its effect.

In Fig. 3, where we apply Eq. $\left( \ref{salsef}\right) $ to pp reactions
in a zero metallicity plasma, we can easily see that electron degeneracy
effects always reduce Salpeter's SEF to a degree which varies according to
the degree of degeneracy (i.e. the degeneracy parameter). For small
densities and high temperatures (e.g. the sun) the effect is not very
important but in more degenerate gases, as one would expect, there is a
notable reduction of the energy production rate.

\section{Nonlinear screening effects}

If the DH potential $\Phi _{DH}\,\,,$ which was derived via Ineqs $\left( 
\ref{condwes}\right) ,\,$ is to be used inside the tunnelling region of the
usual WKB integral that appears in the penetration factor $\Gamma \left(
E\right) $:

\begin{equation}
\Gamma \left( E\right) =\exp \left( \frac{-2}{\hbar }\int_{R}^{r_{tp}}~~[2%
\mu (\Phi _{DH}(r)-E)]^{1/2}dr\right)  \label{wkbint}
\end{equation}
then Ineqs $\left( \ref{condwes}\right) $ should also be valid for the DH
potential throughout that region:

\begin{equation}
\,\frac{Z_{i}e\Phi _{DH}\left( r\right) }{kT}\ll 1\Leftrightarrow \frac{%
Z_{i}Z_{0}e^{2}}{kTr}\exp \left( -\frac{r}{R_{D}}\right) \ll 1
\label{corlim}
\end{equation}
where the index $i$ spans the entire ionic profile of the plasma.

Setting $x=r/R_{D}$ we arrive at 
\begin{equation}
\frac{e^{-x}}{x}\ll \left( \ln f_{\max }\right) ^{-1}  \label{ineqx}
\end{equation}
where $f_{\max }=\exp \left( \frac{Z_{\max }Z_{0}e^{2}}{R_{D}kT}\right) $ is
Salpeter's SEF for the heaviest ion in the plasma interacting with our
central nucleus $Z_{0}e$ $.$

Ineq.$\left( \ref{ineqx}\right) $ provides us with the relative distances
between the reacting nuclei where Eq. $\left( \ref{potdh}\right) $ can be
used. These distances will obviously be fractions of the DH radius and since 
$f_{s}^{\max }>1$ Ineq.$\left( \ref{ineqx}\right) $ will impose the
condition $x>>x_{\min }$. That means that there is critical value for $x$
below which the DH formalism is not valid, or conversely that there is
always a value $x_{0}$ and a parameter $\beta <<1$ so that at all distances
larger than $x_{0}$ the weak screening condition (and for that reason the
DH\ model) is adequately satisfied. If we fix the desired accuracy(i.e. $%
\beta \,$) then $x_{0}$ can be obtained through the following tuning equation

\begin{equation}
\frac{e^{-x_{0}}}{x_{0}}=\beta \left( \ln f_{s}^{\max }\right) ^{-1}
\label{x0beta}
\end{equation}

It is obvious that if we set $\beta =0.1$ (see also Ref. \cite{dzitko}),
then in the linearization of Eq.$\left( \ref{boltzmann}\right) \,$we make an
error of $1\%$ which is acceptable.

Likewise, we can select various values for $\beta $, according to the
particular thermonuclear reaction at hand and the desired precision, thus
tuning and gauging the nonlinear screening effects disregarded by both the S
and M models. In our model we adopt the value of $\beta =0.1$ which
obviously provides a very accurate approximation.

A\ vivid impression of how difficult it is for Ineq. $\left( \ref{corlim}%
\right) $ to be obeyed during the tunnelling process can be obtained by the
following argument: The tunnelling region $R<r<r_{tp}$ \thinspace (see Eq. $%
\left( \ref{wkbint}\right) )\,$\thinspace is defined by the closest distance
between the reacting nuclei $R=R_{1}+R_{2}=1.3\left(
A_{1}^{1/3}+A_{2}^{1/3}\right) fm$ and the classical turning point given by

\begin{equation}
\frac{e^{-x_{tp}}}{x_{tp}}=\frac{E_{0}}{kT\ln f_{s}}  \label{xtp}
\end{equation}
where $E_{0}$ is the most effective energy of the binary reaction $%
_{Z_{1}}^{A_{1}}X+_{Z_{0}}^{A_{0}}X$ given for screened reactions by \cite
{lioliosprc2000}:

\begin{equation}
E_{0}=1.22\left( Z_{1}^{2}Z_{0}^{2}AT_{6}\right) ^{1/3}\xi ^{2/3}\left(
x^{*}\right) ,\,\,\,\,\,\,\xi \left( x^{*}\right) \simeq e^{-x^{*}}\left( 1+%
\frac{x^{*}}{2}+\frac{x^{*2}}{16}\right) ,\text{ }x^{*}=\frac{r_{tp}}{R_{D}}
\end{equation}

Therefore if we really want Ineq. $\left( \ref{corlim}\right) $ to be obeyed
throughout the tunnelling region then that should be the case at touching
distances as well , that is:

\begin{equation}
\frac{Z_{\max }Z_{0}e^{2}}{1.3\left( A_{0}^{1/3}+A_{\max }^{1/3}\right) }%
\exp \left( -\frac{1.3\left( A_{0}^{1/3}+A_{\max }^{1/3}\right) }{R_{D}}%
\right) \ll kT
\end{equation}
or else

\begin{equation}
T_{6}\gg \frac{12880Z_{0}Z_{\max }}{\left( A_{0}^{1/3}+A_{\max
}^{1/3}\right) }
\end{equation}
If we assume, for example, the pp reaction in the same solar region as
before then the WS condition is always obeyed if $T_{6}\gg 29274$ which is a
temperature too high even for supernova explosions. This last condition
proves that the weak-screening condition is hardly ever obeyed across the
entire Coulomb barrier under any circumstances..

As an example, let us consider a main sequence star where the proton-proton
\thinspace $\left( pp\right) $ reaction takes place (e.g. the center of the
sun). There we have $R_{D}=24819\,fm$ and of course $f_{s}^{pp}=1.049,%
\,E_{0}=5.6\,keV\,,kT=1.2\,keV.\,$The classical turning point derived from
Eq. $\left( \ref{xtp}\right) $ is then $r_{tp}=10^{-2}R_{D}$ and the
touching distance is of the order of a few fermis, as usual. Let us assume
for simplicity that $Z_{\max }=2$ (that of helium) which means that $f_{\max
}=1.10.$ Then Ineq. $\left( \ref{ineqx}\right) $ shows that potential $%
\left( \ref{potdh}\right) $ is only valid when $r>>8.\,7339\times
10^{-2}R_{D}.$ However, the penetration factor $\Gamma \left( E\right) $
that appears in the thermonuclear reaction rate formalism and in fact
derives Eq. $\left( \ref{fs}\right) ,$ involves an integral running from $R$
to $r_{tp}$. In other words the linearization we performed in Sec. III.
(especially for ions with higher atomic numbers) forbids the use of Eq. $%
\left( \ref{potdh}\right) $ in the tunnelling region, and the subsequent
derivation of Salpeter's formula. This is due to the fact that the DH\ model
was invented to handle global properties between ions and electrons in
electrolytes which interact at far larger distances than the ones at which
thermonuclear reactions take place.

A crucial question is whether there is any interference between the
degeneracy conditions and the WS one. Even in CD environments the screening
condition for ions is the same as in ND ones. For electrons in PD gases,
however, it reads:

\begin{equation}
\frac{e\Phi }{kT}<<-a
\end{equation}
which is apparently less strict than in ND electron gases.\thinspace For
example, in ND gases the ratio $\left( e\Phi /kT\right) $ must be much
smaller than in PD ones thus limiting further the actual radius of validity
of the screened Coulomb potential.\thinspace In (at least) SD gases the
strongest screening condition remains that of ions, namely Ineq. $\left( \ref
{ineqx}\right) $. This is just as expected because in dense, (relatively)
cold gases, electrons play a minor role as the provide a (relatively)
uniform negative background where ionic screening plays the major role, as
it leads to a crystal-like configuration.

\section{A new model for screened thermonuclear reactions}

Most authors who have dealt with screening corrections devise a mean field
screened Coulomb potential (see Ref.\cite{lioliostf2001} and references
therein)

\begin{equation}
\Phi \left( r\right) =\frac{Z_{0}e}{r}f\left( \frac{r}{R_{s}}\right)
\end{equation}
where $R_{s}$ is the screening radius which depends on plasma
characteristics and is of course a function of its density and temperature.
The potential energy of the interaction between nuclei $\left(
Z_{0}e,\,Z_{1}e\right) $ is $\left( Z_{1}e\right) \Phi \left( r\right) $
which, in turn, is used in the evaluation of the action integral that
appears in the WKB treatment of the thermonuclear reaction rate. As the
classical turning point lies deeply inside the screening cloud, the usual
approach drops higher terms appearing in the integral and by a method, which
is now well known\cite{lioliosprc2001}, there appears an exponential
corrective term widely known as the screening enhancement factor (SEF).
Similar methods have been used in the treatment of screening phenomena in
laboratory astrophysical reactions\cite{lioliostf2001}. A very common error
in such studies\cite{cameron,harisson} is disregarding the binary nature of
the reaction rate $r_{ij}$ between nuclei $i,j$, which demands that its
formula should be commutative that is $i,j$ should be interchangeable. This
demand forced Mitler to use Helmholtz free energies in his calculations
instead of the usual energy shift which was erroneously used by other authors%
\cite{cameron,harisson}. Mitler's method, however, has made the arbitrary
assumption that at close distances from all nuclei in the plasma the charge
density is practically equal to the average electron density in the plasma.
He then assumed that at a certain distance from the target nucleus the
charge density is equal to the DH one, normalized appropriately.

We can now clearly state our two major objections to Mitler's model:

First, although he presents some proof that near the origin the actual
density is slightly higher than the average electron density in the plasma,
his rigid core assumption is obviously unnatural. In screened plasmas the
density is supposed to follow a natural reduction pattern which at large
distances should coincide with the DH one.

Second, he calculates his critical distance modifying the DH density which
is well established at large distances. It is obvious from the previous
analysis that beyond a certain point calculated through Eq. $\left( \ref
{x0beta}\right) $ the DH density (and the respective interaction energy) is
an excellent approximation.

It is, therefore, much more natural to inverse Mitler's method and adopt the
DH\ formalism beyond point $x_{0}$ so that we can calculate the internal
charge density. To that end we assumed a very natural behavior for the
charge density around the nucleus and devised the model that follows:

\subsection{Evaluation of the critical distance and the central density}

Let us assume that the charge density $\rho _{in}\left( r\right) $ around
the nucleus $Z_{0}e$, instead of being constant, is an exponentially
decreasing function of distance of the form

\begin{equation}
\rho _{in}\left( r\right) =\rho \left( 0\right) \exp \left( -r/R_{0}\right)
\label{newden}
\end{equation}
At the critical point $r_{0}$ the density assumes the usual DH\ form that is

\begin{equation}
\rho _{in}\left( r_{0}\right) =\rho _{DH}\left( r_{0}\right) =\left( -\frac{%
Z_{0}e}{4\pi R_{D}^{2}}\right) \frac{\exp \left( -r_{0}/R_{D}\right) }{r_{0}}
\end{equation}
The critical point will of course be calculated using Eq. $\left( \ref
{x0beta}\right) ,$ which was thoroughly analyzed in the previous section
while the central charge density will be

\begin{equation}
\rho \left( 0\right) =\rho _{DH}\left( x_{0}\right) \exp \left(
x_{0}^{^{\prime }}\right)  \label{deneq}
\end{equation}
where we have set $x=r/R_{D},x_{0}=r_{0}/R_{D}$ $,x^{^{\prime
}}=r/R_{0},\,x_{0}^{^{\prime }}=r_{0}/R_{0}.$

The DH charge density at $r=R_{D}$ is $\rho _{DH}\left( R_{D}\right)
=-\left( e^{*}\right) ^{-1}\left( Z_{0}e/4\pi R_{D}^{3}\right) \,$, while
the densities at $x_{0}$ and $R_{D}$ are related through

\begin{equation}
\rho _{in}\left( x_{0}\right) =\rho _{DH}\left( x_{0}\right) =\frac{%
e^{1-x_{0}}}{x_{0}}\rho _{DH}\left( R_{D}\right)
\end{equation}
(note that we set $e^{*}=\exp \left( 1\right) \,$to avoid confusion with the
electron charge when deemed necessary). Therefore using Eq. $\left( \ref
{deneq}\right) $ the central density will be

\begin{equation}
\rho \left( 0\right) =\frac{e^{1+x_{0}^{^{\prime }}-x_{0}}}{x_{0}}\rho
_{DH}\left( R_{D}\right) =\frac{e^{x_{0}^{^{\prime }}-x_{0}}}{x_{0}}\left( -%
\frac{Z_{0}e}{4\pi R_{D}^{3}}\right)  \label{centralden}
\end{equation}
The parameter$\,x_{0}^{^{\prime }}$ can be calculated using the charge
normalization condition

\begin{equation}
\int_{V}\rho \left( r\right) dV=-Z_{0}e  \label{chnorm}
\end{equation}
The meaning of that condition is that all the charge surrounding the nucleus 
$Z_{0}e$ will be of course the total plasma charge (i.e. zero) minus the
positive ion itself which of course is missing from the total plasma as it
is located in the center of the configuration. Therefore the total charge of
the configuration will be negatively charged by the quantity ($-Z_{0}e).$

Eq. $\left( \ref{chnorm}\right) $ can now be written

\begin{equation}
\int_{0}^{r_{0}}\rho _{in}\left( r\right) 4\pi r^{2}dr+\int_{r_{0}}^{\infty
}\rho _{DH}\left( r\right) 4\pi r^{2}dr=-Z_{0}e
\end{equation}

Inserting the definition of the densities we have

\begin{equation}
4\pi \rho \left( 0\right) R_{0}^{3}\int_{0}^{x_{0}^{^{\prime
}}}x^{2}e^{-x}dx-Z_{0}e\int_{x_{0}}^{\infty }xe^{-x}dx=-Z_{0}e
\label{central}
\end{equation}
The integrals that appear in the previous expression are trivial and can be
calculated analytically. Therefore, if we set

\begin{equation}
I_{1}\left( x_{0}^{^{\prime }}\right) =\int_{0}^{x_{0}^{^{\prime
}}}x^{2}e^{-x}dx=-e^{-x_{0}^{^{\prime }}}\left( x_{0}^{^{\prime
}2}+2x_{0}^{^{\prime }}+2\right) +2  \label{i1}
\end{equation}
and

\begin{equation}
I_{2}\left( x_{0}\right) =\int_{x_{0}}^{\infty }xe^{-x}dx=e^{-x_{0}}\left(
x_{0}+1\right)  \label{i2}
\end{equation}
then Eq. $\left( \ref{central}\right) $ reads

\begin{equation}
4\pi \rho \left( 0\right) R_{0}^{3}I_{1}\left( x_{0}^{^{\prime }}\right)
-Z_{0}eI_{2}\left( x_{0}\right) =-Z_{0}e  \label{i1i2}
\end{equation}
Combining Eq.$\left( \ref{deneq}\right) ,\left( \ref{i1}\right) ,\left( \ref
{i2}\right) $ and Eq.$\left( \ref{i1i2}\right) $ we obtain

\begin{equation}
\frac{e^{x_{0}^{^{\prime }}}}{x_{0}^{^{\prime }3}}I_{1}\left(
x_{0}^{^{\prime }}\right) =\frac{e^{x_{0}}}{x_{0}^{2}}\left[ 1-I_{2}\left(
x_{0}\right) \right]  \label{final1}
\end{equation}
Since $R_{0}=\left( x_{0}/x_{0}^{^{\prime }}\right) R_{D},$ Eq.$\left( \ref
{final1}\right) $ and Eq. $\left( \ref{x0beta}\right) $ give all the
information we want for our model.

\subsection{Derivation of the screened Coulomb potential}

Once the inner charge density has been derived, Poisson's equation can
produce the screened Coulomb potential around the reacting nuclei:

\begin{equation}
\nabla^{2}\Phi =-4\pi \rho \left( 0\right) \exp \left( -r/R_{0}\right)
\label{poissonap}
\end{equation}
obeying the boundary conditions

\begin{equation}
\lim_{r\rightarrow 0}\Phi \left( r\right) =\frac{Z_{0}e}{r}%
,\,\,\,\lim_{r\rightarrow r_{0}}\Phi \left( r\right) =\Phi _{DH}\left(
r\right)  \label{bcond1}
\end{equation}
The solution is of the form:

\begin{equation}
\Phi \left( r\right) =\frac{Z_{0}e}{r}F\left( r\right)  \label{ff}
\end{equation}
where $F\left( r\right) \,$obeys the boundary conditions

\begin{equation}
F\left( 0\right) =1,\,\,\,F\left( r_{0}\right) =\exp \left(
-r_{0}/R_{D}\right)  \label{bcond2}
\end{equation}
Having assumed spherical symmetry for the screening effect Eq.$\left( \ref
{poissonap}\right) $ can be written

\begin{equation}
\frac{d^{2}\Phi }{dr^{2}}+\frac{2}{r}\frac{d\Phi }{dr}=-4\pi \rho \left(
0\right) \exp \left( -r/R_{0}\right)  \label{poisradial}
\end{equation}
Combining Eq. $\left( \ref{ff}\right) $, and Eq. $\left( \ref{poisradial}%
\right) $ we can derive the general form of the screened Coulomb potentials
which correspond to our charge density distribution, that is

\begin{equation}
F\left( r\right) =c_{2}+c_{1}r-\frac{4\pi \rho \left( 0\right) R_{0}^{3}}{%
Z_{0}e}\left( 2+\frac{r}{R_{0}}\right) e^{-r/R_{0}}
\end{equation}
Using the respective boundary conditions given by Eqs. $\left( \ref{bcond2}%
\right) $ and after some tedious algebra the constants $c_{2},c_{1}$ are:

\begin{equation}
c_{2}=1-2\frac{x_{0}^{2}}{x_{0}^{^{\prime }3}}e^{x_{0}^{^{\prime }}-x_{0}}
\end{equation}
and

\begin{equation}
c_{1}=-R_{D}^{-1}\left[ \frac{1}{x_{0}}-\frac{e^{-x_{0}}}{x_{0}}-2\frac{x_{0}%
}{x_{0}^{^{\prime }3}}e^{x_{0}^{^{\prime }}-x_{0}}+\frac{x_{0}}{%
x_{0}^{^{\prime }3}}\left( 2+x_{0}^{^{\prime }}\right) e^{-x_{0}}\right]
\end{equation}

\subsection{Derivation of the screening enhancement factor (SEF)}

Having obtained both the charge density and the associated screened Coulomb
potential the screening energy can be obtained in the following way: As
noted before, Mitler makes use of the free energies of the cloud-ion
configuration. Although Mitler's constant density allows the derivation of
an analytic formula for the screening energy the author's model would
involve a lot more work. Therefore, in the present paper we will only obtain
SEFs for some special cases, which, however, cover a wide spectrum of
astrophysical reactions:

a) The quantity $Z_{1}e\Phi \left( r\right) $ is symmetric in 1 and 2 and
the impinging nucleus $Z_{1}e$ is considered unscreened, which means that it
carries no screening cloud as it collides with ion $Z_{0}e$ which creates
the potential $\Phi \left( r\right) .$

b) The quantity $Z_{1}e\Phi \left( r\right) $ is not symmetric in 1 and 2
but the charge $Z_{0}e$ is considerably larger than $Z_{1}e$ so that we can
safely disregard the cloud associated with nucleus $Z_{1}e$ which can be
considered unscreened. In such a case, although the quantity $Z_{1}e\Phi
\left( r\right) \,$is not symmetric, the classical turning point lies so
deeply inside the screening configuration that it makes no difference
whether the cloud is attributed to $Z_{0}e$ or $Z_{1}e$.

If the electron speed was much higher than that of the reacting nuclei as
suggested in Ref. \cite{dzitko}, then we could take into account the cloud
perturbation induced by the projectile by setting $Z_{0}\rightarrow \left(
Z_{0}+1\right) ,$ since in that case the cloud would respond fast enough.
However, this is not the case since we should not compare the electron speed
to the maximum thermal energy of ions as in Ref.\cite{dzitko}, but instead
we should compare the electron speed to the most effective energy of nuclear
interaction. As is well known the latter is larger than the former in
non-relativistic gases. We solve that problem by assuming that there are two
limits: The sudden limit (SD) where we assume that the cloud remains
perfectly rigid during tunnelling and the adiabatic limit (AD) where we
assume that the cloud responds so fast that during tunnelling the source of
polarization is not $Z_{0}$ but $Z_{0}+1,$ instead. Therefore the central
charge to be used will be $Z_{0}^{*}e$ so that:

\begin{equation}
Z_{0}<Z_{0}^{*}<Z_{0}+1
\end{equation}
$\,$

Obviously for heavily charged nuclei (i.e. $Z_{0}\gg 1$ )\thinspace
\thinspace both limits yield practically the same result.

Since the classical turning point for a particular reaction is always much
smaller than $R_{0}$ the screened Coulomb potential to be used across the
barrier is truncated so that

\begin{equation}
\Phi \left( r\right) =\frac{Z_{0}e}{r}-\frac{Z_{0}e}{R_{D}}G\left(
x_{0},x_{0}^{^{\prime }}\right) +O\left( r^{2}\right)
\end{equation}
where the quantity $G\left( x_{0},x_{0}^{^{\prime }}\right) $ is

\begin{equation}
G\left( x_{0},x_{0}^{^{\prime }}\right) =\frac{1}{x_{0}}-\frac{e^{-x_{0}}}{%
x_{0}}-2\frac{x_{0}}{x_{0}^{^{\prime }3}}e^{x_{0}^{^{\prime }}-x_{0}}+\frac{%
x_{0}}{x_{0}^{^{\prime }3}}\left( 2+x_{0}^{^{\prime }}\right) e^{-x_{0}}+%
\frac{x_{0}}{x_{0}^{^{\prime }2}}e^{x_{0}^{^{\prime }}-x_{0}}  \label{gxx}
\end{equation}

In the above two cases, following the usual mechanism\cite{lioliosprc2000},
we derive a shift in the relative energy (the screening energy)

\begin{equation}
U_{e}=\frac{Z_{0}Z_{1}e^{2}}{R_{D}}G\left( x_{0},x_{0}^{^{\prime }}\right)
\end{equation}
with the modification $Z_{0}\rightarrow Z_{0}^{*}$ for protons impinging on
heavily charged nuclei (e.g. reactions of the rp process.

As we know the DH screening energy is $U_{DH}=Z_{0}Z_{0}e^{2}R_{D}^{-1}$ and
the respective SEF is given by Eq.$\left( \ref{fs}\right) .$ Thus, if we use
the definition of $G\left( x_{0},x_{0}^{^{\prime }}\right) $ the screening
energy of the new model will be

\begin{equation}
U_{e}=U_{DH}G\left( x_{0},x_{0}^{^{\prime }}\right)  \label{uu}
\end{equation}
while using the definition of Salpeter's $f_{s}$ the new improved SEF will
be:

\begin{equation}
f=f_{s}^{G\left( x_{0},x_{0}^{^{\prime }}\right) }  \label{mysef}
\end{equation}
Eq. $\left( \ref{mysef}\right) $ is the formula that should replace
Salpeter's in all cases while Mitler's is replaced only in astrophysical
reaction which satisfy the two above mentioned conditions.

\section{Application of the new model to the solar pp reaction and the
associated SNF}

Let us apply our model to the pp reaction is the sun and see how it modifies
the SNF. In the solar plasma we can safely assume that the heaviest metal
which exists in non-negligible quantities and can have a (very small indeed)
effect on the DH$\;$radius is oxygen that burns last in the CNO cycle so
that $Z_{\max }=8.$ All other metals with $Z>8$ can be totally disregarded
in the plasma. We will apply our model in the solar region of maximum energy
production that is\cite{bahcallbook} at a distance of $0.075$ solar radii
where $T_{6}=14$ and $\rho =105\,g/cm^{3}.$ At such a distance the most
important isotopic abundances\cite{bahcallbook} (the fractions by mass of
hydrogen $\left( X\right) $ and helium $\left( Y\right) $) are $X=0.48$ and $%
Y=0.5.\,$Then $Z_{\max }=8,$ and of course $f_{S}^{\max }\left( Z_{\max
}=8\right) =1.52\,\,$so that Eq.$\left( \ref{x0beta}\right) $ reads $\exp
\left( -x_{0}\right) =0.237x_{0}$ with a solution of $x_{0}=1.\,1781$. Using
Eq. $\left( \ref{final1}\right) $ we obtain $x_{0}^{^{\prime }}=2.716.\,$%
Therefore the critical point is $r_{0}=1.1781R_{D}\,$while the inner
screening radii will now be $R_{0}=0.433R_{D.}$ Thus, the DH model cannot be
used at distances shorter than $r=1.178R_{D}$ without committing an error.
On the other hand the central charge density will be given by Eq. $\left( 
\ref{centralden}\right) \,$so that $\rho \left( 0\right) \simeq 10\rho
_{DH}\left( R_{D}\right) $ $.$ Using Eq. $\left( \ref{uu}\right) $ we see
that our screening energy is $U_{e}/U_{DH}=0.87.\,\,$If instead we had
assumed that $Z_{\max }=2$, as we did before, the critical point would have
been $x_{0}=0.\,55129$ while at the same time $x_{0}^{^{\prime }}=2.0.$ Then
the central density would be $\rho \left( 0\right) \simeq 134\rho
_{DH}\left( R_{D}\right) \,$and the screening energy $U_{e}/U_{DH}=0.93\,$%
\thinspace (The same effects would be observed if $\,$we set $\beta =0.4$
and retain the value of $Z_{\max }=8).$ Note that the smallest possible
value of $x_{0}$ that can be derived via the present model at solar
conditions is obtained for a purely hydrogen (OCP) plasma $\left( \rho
=105,T_{6}=14\right) $ by abusing Ineq.$\left( \ref{ineqx}\right) $ to its
limits (i.e. by setting $\beta =1$,$\,Z_{\max }=1).$ In such case we have $%
x_{0}=0.0\,6867\,$which yields through Eq. $\left( \ref{final1}\right) $ $%
x_{0}^{^{\prime }}=1.52$ $.\,$These values yield a central density of $\rho
\left( 0\right) =170\rho _{DH}\left( R_{D}\right) $ and a screening energy
of $U_{e}/U_{DH}=0.99.\,$For the only possible thermonuclear reaction in
such a plasma (i.e. hydrogen) we have a classical turning point $%
x_{tp}=0.036 $ and a DH radius $R_{D}=16261fm.$ In fact we see that, for all
reactions, as $x_{0}\rightarrow 0,$ then $x_{0}^{^{\prime }}\rightarrow 1.45$
and $U_{e}/U_{DH}\rightarrow 1$ that is, as expected, we derive the same S
model screening energy and of course the same infinite central density.

A very important argument is that the accuracy of our model is greatly
improved if we are able to make assumptions for the stellar composition
which reduce the distance $\left( x_{0}-x_{tp}\right) $, thus reducing the
uncertainty across the barrier. If the plasma composition allowed the
penetration of $x_{0}$ deeply into the tunnelling region (i.e. $%
x_{0}<x_{tp}, $ which is not possible for normal stellar environments$)$
then we could really achieve maximum precision. This is due to the fact that
the DH\ density is well established for $x>x_{0}\,$while Eq. $\left( \ref
{newden}\right) \,\,$used at $x<x_{0}\,$is artificial and depends heavily on
the tuning accuracy that we adopt each time.

According to the above calculations for the pp reaction in the sun the SEF
value obtained through the present model is given as a function of
Salpeter's SEF

\begin{equation}
f^{pp}=\left( f_{s}^{pp}\right) ^{q}
\end{equation}
where $0.87<q<0.93$ so that

\begin{equation}
1.\,042<f^{pp}<1.045.  \label{ineqpp}
\end{equation}

Let us now compare our results to Mitler's: The critical distance where,
according to the M model, the charge density abruptly ceases to be equal to
the average electron density and assumes a DH\ form is roughly $%
r_{0}=0.1R_{D}.$ (i.e. $x_{0}^{M}=0.1$ ). We can easily observe that, if we
follow the M model, then across the tunnelling region which starts at the
classical turning point ($r_{tp}=0.01R_{D},$ see Sec. VI.) the collision
takes place through a rigid electronic cloud barren of any ions or any kind
of charge density variation. According to the M formulas we obtain a
screening shift which is $95\%$ (or $92\%$ if the second proton is also
screened) of that obtained through the S model. Thus for the $pp$ reaction
we have $f_{Mit}^{pp}=1.046\,$(or $1.045$ for both protons screened-see also
Ref.\cite{ricci}), which coincides with the most probable SEF of our model.
Note that it matters very little if we consider both protons screened or
only one, just as we assumed in our model. We see that although both
Mitler's model and the present one give a smaller SEF than the S model there
is an important difference between them. Actually, in order to obtain
Mitler's critical distance we should tune our weak-screening accuracy so
that our $x_{0}$ coincides with $x_{0}^{M}.$ If we assume that $Z_{\max
}=2\,\,$\thinspace the value $x_{0}=x_{0}^{M}$ can only be obtained by
setting $\beta =0.8$ $\,\,$in Eq. $\left( \ref{x0beta}\right) $ which is an
extremely insufficient weak-screening assumption as it abuses Ineq.$\left( 
\ref{condwes}\right) $ to its limits. As regards the constant density
assumed by Mitler we see that when $Z_{\max }=2$ \thinspace our model gives
for pp reactions\thinspace $\rho _{in}\left( x_{0}\right) =0.13\rho \left(
0\right) $, $\rho _{in}\left( x_{tp}\right) =0.96\rho \left( 0\right) $ $\,$%
and $\,\rho _{in}\left( x_{0}^{M}\right) =0.69\rho \left( 0\right) $,
\thinspace which means that Mitler's assumption is not justified for the
distances he claims (i.e. $x_{0}^{M}=0.1)$ but it is perfectly justified
across the tunnelling region. Moreover, we need to underline that the actual
value of the central density $\rho \left( 0\right) $ is immaterial to the
novel model as it just an artificial one and its only purpose is to derive
the correct SEF. Mitler's model, on the other hand, does exhibit a very weak
dependence on the actual value of the central density\cite{ricci} which for
solar conditions can be totally disregarded. Owing to the last two aspects
of compatibility between the two models it is plausible that they should
yield roughly the same SEFs, as is actually the case.

Therefore the most accurate value of the pp SEF in the sun is of course the
upper limit of the double inequality$\,\left( \ref{ineqpp}\right) $ with an
uncertainty which obviously lies within its upper and lower limit. If
instead we had assumed that both protons are screened then that would have
yielded a negligibly smaller SEF.

We can now study the uncertainties induced on the solar neutrino fluxes
(SNF) by the present improvements of Salpeter's and Mitler's models
employing the proportionality formulas\cite{ricci,lioliosprc2000} which
relate the screened (SC) and unscreened (NOS)\ SNF. In these formulas we
assume that all reactions are unscreened except for the pp one in order to
isolate the effect for the most important solar reaction. In such a case 
\begin{equation}
\frac{\Phi _{pp}^{SC}}{\Phi _{pp}^{NOS}}=\left( f_{s}^{pp}\right) ^{0.14q},%
\frac{\Phi _{Be}^{SC}}{\Phi _{Be}^{NOS}}=\left( f_{s}^{pp}\right) ^{-1.25q},%
\frac{\Phi _{B}^{SC}}{\Phi _{B}^{NOS}}=\left( f_{s}^{pp}\right) ^{-2.95q},%
\frac{\Phi _{N,O}^{SC}}{\Phi _{N,O}^{NOS}}=\left( f_{s}^{pp}\right) ^{-2.75q}
\end{equation}

In Fig.4. we can observe the dependence of the solar neutrino fluxes (SNF)
on the (uncertainty of the) proton-proton SEF $f^{pp}$. The ratio of the
screened SNF $\left( \Phi ^{SC}\right) \,$versus the unscreened ones $\left(
\Phi ^{NOS}\right) \,$are given for various neutrino producing solar
reactions with respect to the $q$ quantity defined in the text. The most
probable values given by the new improved model are those given at $q=0.93,$
while the values corresponding to Salpeter's formula are those at $q=1.$ The
vertical bars at $q=0.87$ and $q=0.93$ represents the boundaries of
reasonable assumptions for the solar composition and the WS limit. To
validate the region lying outside those barriers of uncertainty either we
have to assume that the solar plasma has a different composition that the
standard model predicts or to compromise our weak screening linearization by
tuning down the accuracy so that $\beta >0.1$. Note that if we attempt to
tighten the WS condition assuming that $\beta <0.1,$ then we actually push
the critical point $x_{0}$ further from the classical turning point which
reduces the accuracy of the model.

Admittedly, there is some interference between the screening uncertainties
of all neutrino producing reactions whose effects sometimes cancel each other%
\cite{ricci}. Other authors\cite{gruzinov} evaluate the total uncertainty by
simply using the S model. We might as well follow their example and use the
model derived in this paper which is more accurate than the S one anyway.
However that approach: a) would cloud the effect of the most important solar
reaction (i.e. pp) for which our model is legitimate and b)\ would abuse our
model (just at it abuses the S one) since both conditions (a,b) of Sec. IV.
C would be violated.

The compatibility of our model to the measurable macroscopic solar
quantities is well established. For example, the enhancement of the energy
production due to screening is subject to a rigid constraint imposed by the
experimentally measured solar luminosity which must be kept constant,
regardless of the solar model selected each time. Thus, the ratio of the
screened temperature $\left( T^{SC}\right) \,$versus the unscreened one $%
\left( T^{NOS}\right) \,$is given by the quantity\cite{ricci}\thinspace $%
\left( T^{SC}/T^{NOS}\right) =\left( f^{pp}\right) ^{-1/8}.$ On the other
hand the ratio of $\rho /T^{3}\,$is approximately constant along the whole
stellar profile so that\cite{lioliosdeform} $\left( \rho ^{SC}/\rho
^{NOS}\right) =\left( f^{pp}\right) ^{-3/8}.$ It is easy to see that the new
model is perfectly compatible with the density and the temperature of the
standard solar model .

\section{Discussion and conclusions}

In this paper we first derive the ND SEF from the first principles of the
Thomas-Fermi theory whose microscopic nature allows the study of nonlinear
screening effects as opposed to the macroscopic nature of Boltzmann's
statistical theory employed by other authors\cite{salpeter}. In a recent
paper the nonlinear corrections to Salpeter's formula were studied by
numerically solving Poisson's equation\cite{gruzinov}. In the present work
we present a very transparent analytical investigation which allows us to
control the associated error by tuning the weak screening condition at will.

On the other hand, in a PD environment, the S model takes into account
degeneracy effects by the introduction of a degeneracy factor $\theta \left(
a\right) $ whose numerical calculation involves some improper integrals
which delay the computing process in stellar evolution codes. In this paper,
using the principles of the TF\ theory, there is produced, for the first
time to our knowledge, a very handy analytic degeneracy factor, which can be
readily used in the DH radius involved in Salpeter's, Mitler's and the
author's models.

Salpeter's SEF was derived on the assumption that the DH\ model can be used
inside the tunnelling region when calculating the penetration factor in
thermonuclear reaction rates. However, as we have proved there is always an
internuclear distance inside that region where the DH model breaks down
because the assumptions that led to its derivation are no longer valid
(i.e.e Ineqs $\left( \ref{condwes}\right) ).$

Mitler, on the other hand, derived a SEF for all densities by making an
assumption which we prove here to be correct, to some extent, for the weakly
screened solar environment. He assumed that the charge density around ions
in all stellar environments is equal to the average electron density in the
plasma. He then evaluated some critical distance from the ion where the
constant density model is abruptly replaced by the DH model. The alarming
discontinuity that appears at that critical distance is disregarded as is
the unnatural assumption of a rigid electron core around the ion in all
stellar plasmas. Admittedly Mitler devotes a few lines to qualitatively
justify his assumptions but, as the author has already asserted, this effort
doesn't lift their arbitrariness.

A new model is presented here which avoids Mitler's constant electron
density assumption and remedies the break-down of the DH\ model inside the
tunnelling region from which Salpeter's model suffers. Instead of making
Mitler's assumption about the inner density, we have decided to work
backwards. That is we use the TF model to establish the minimum distance
from the ion where the DH\ model is correct and we seek the inner density's
details by making the most natural assumption that the charge density from
the ion to the critical point is an exponentially decreasing function of
distance. This way we have full control of the DH model nonlinearities,
since by tuning the WS condition we can actually take into account the error
we commit in the calculation of screening effects. Note that this novel
model tends to the S one when its nonlinear accuracy is tuned down to a
minimum, which shows vividly the improvement over the M model and especially
of its generator, the S one.

In short the new model should be used as follows: In weakly screened
thermonuclear stellar plasmas consisting mainly of hydrogen and helium we
evaluate $x_{0}$ through Eq. $\left( \ref{x0beta}\right) \,$setting $\beta
=0.1$ and $Z_{\max }=2$ .Then we resort to Eq. $\left( \ref{final1}\right) $
in order to evaluate $x_{0}^{^{\prime }}.\,$The novel SEF, which takes into
account degeneracy and nonlinear screening effects, will be given by Eq. $%
\left( \ref{mysef}\right) $ , which is actually Salpeter's SEF raised to the
power of $G\left( x_{0},x_{0}^{^{\prime }}\right) $ (given by Eq. $\left( 
\ref{gxx}\right) )$

The new model is particularly valid for pp reactions as well as for
reactions of protons with heavily charged nuclei (e.g. the rp process
currently under study by the author). We have applied the improved model to
the solar pp reaction and have shown that the respective non-linear
correction to Salpeter's SEF is at most $\left( \sim 1\%\right) $ , which
coincides with the results of Ref. \cite{gruzinov}. We then isolate the
effect of the novel pp SEF on the solar neutrino fluxes evaluating the
associated uncertainly which is now confined in a robust way within very
reliable limits.

{\bf FIGURE CAPTIONS}

Fig.1. The degeneracy factor as obtained through various approximations: The
solid curve represents Eq. $\left( \ref{dp0}\right) $ while the dotted curve
represents the much simpler formula given by Eq. $\left( \ref{dp1}\right) .$
The dash-dot-dotted curve, which stands apart from all other approximations,
represents Eq.$\left( \ref{dp2}\right) .$ The other curves represent
formulas which are in use nowadays, namely the dashed curve stands for
Mitler's guessed formula given by Eq. $\left( \ref{dpmit}\right) $ and the
dash-dotted one is the limit for CD stellar environments given by Eq. $%
\left( \ref{dpfermi}\right) $. The three vertical bars stand for three
different limits. The first one (labeled as 1) signifies the boundary of the
present model. To the right of that barrier Eq. $\left( \ref{dp1}\right) $
is no longer valid. The second one (labeled as 2) represents the SD limit as
defined by Ineq. $\left( \ref{ineqsd}\right) ,$ while the third one (labeled
as 3) is the threshold beyond which the gas is in a CD state and the use of
Eq. $\left( \ref{dpfermi}\right) $ is perfectly legitimate. The region
labeled as ID\ (intermediate degeneracy) is defined by a vertical bar at $%
a=-2\,$\thinspace (not shown) the degeneracy parameter at which the ND and
the CD electron gas pressures are equal.

Fig. 2. Degeneracy zones of the equation of state of an electron gas. The
solid line ($a=2$) separates the ND from the WD regime. The dashed line $%
(a=-2)$ stands for the ID\ limit while the dotted line $(a=-5)$ stands for
the SD limit. Beyond the dash-dotted line $(a=-10)$ the gas is in a CD\
state. The solid lines labeled as $H-H$ and $H-He$ refer to the wrong WS
limit (expressed by Ineq. $\left( \ref{misi}\right) )\,$for a screened
proton in a zero-metallicity plasma. Various stellar environments are shown
in the map such as that of a neutron star, or the ones encountered during
explosive hydrogen burning and x-ray bursts.

Fig.3. Salpeter's SEF for various stellar environments (zero metallicity
scenario) with respect to density when the temperature is kept constant. By
assuming that $T_{6}=10$ ($T_{6}=5)\,$the solid (the dotted) and the dashed
(dash-dotted) curves are derived where the former takes into account the
electron degeneracy effect while the latter assumes that degeneracy is
turned off. The negative numbers attached to the curves labeled as ''ON''
represent the values of the degeneracy parameter $a$ described in the text.

Fig.4. The dependence of the solar neutrino fluxes (SNF) on the (uncertainty
of the) proton-proton SEF $f^{pp}$. The ratio of the screened SNF $\left(
\Phi ^{SC}\right) \,$versus the unscreened ones $\left( \Phi ^{NOS}\right)
\, $are given for various neutrino producing solar reactions with respect to
the $q$ quantity defined in the text. The most probable values given by the
new improved model are those given at $q=0.93,$ while the values
corresponding to Salpeter's formula are those at $q=1.$ The vertical bars
represents the boundaries of reasonable assumptions for the solar
composition and the WS limit.

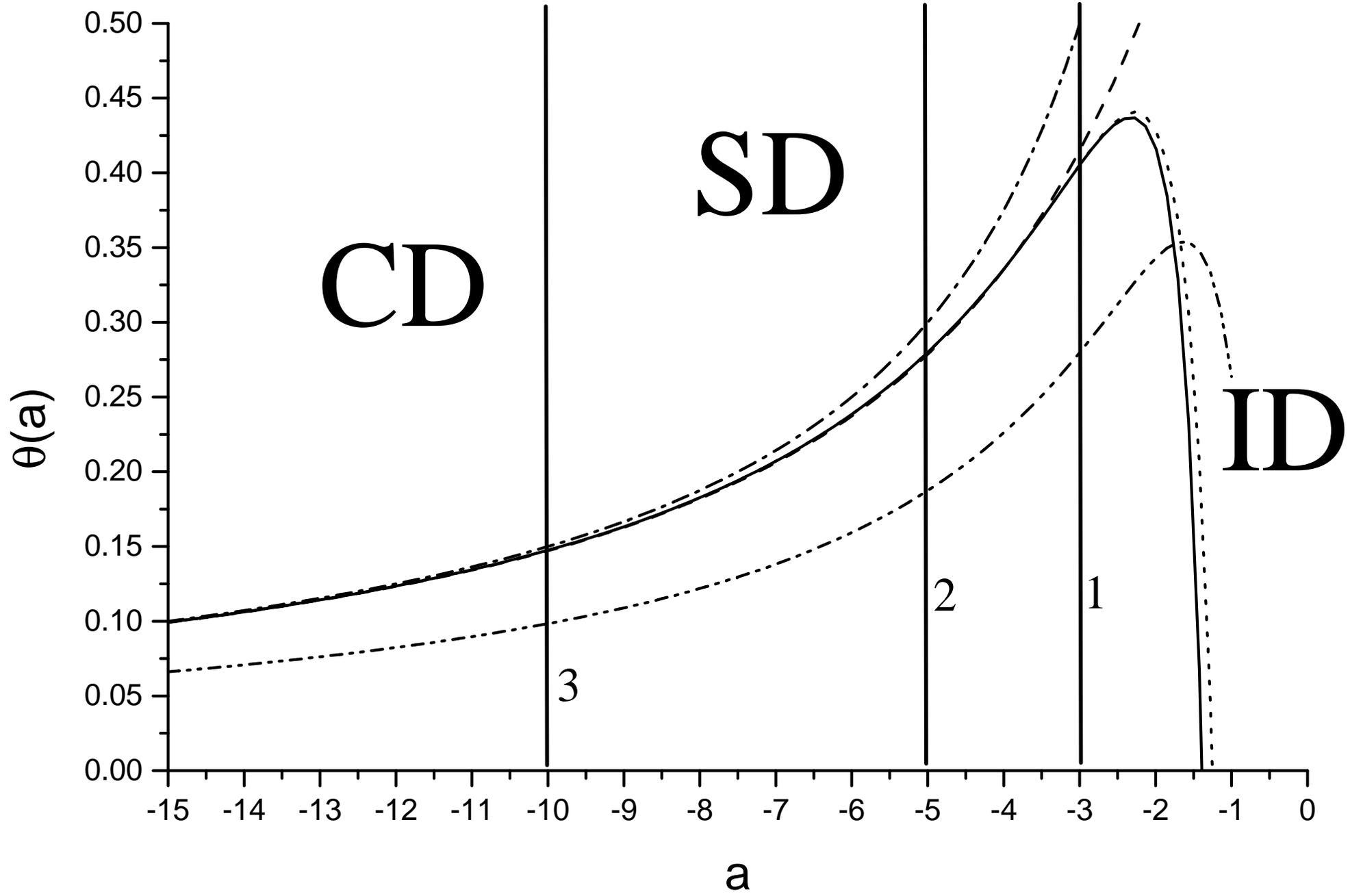
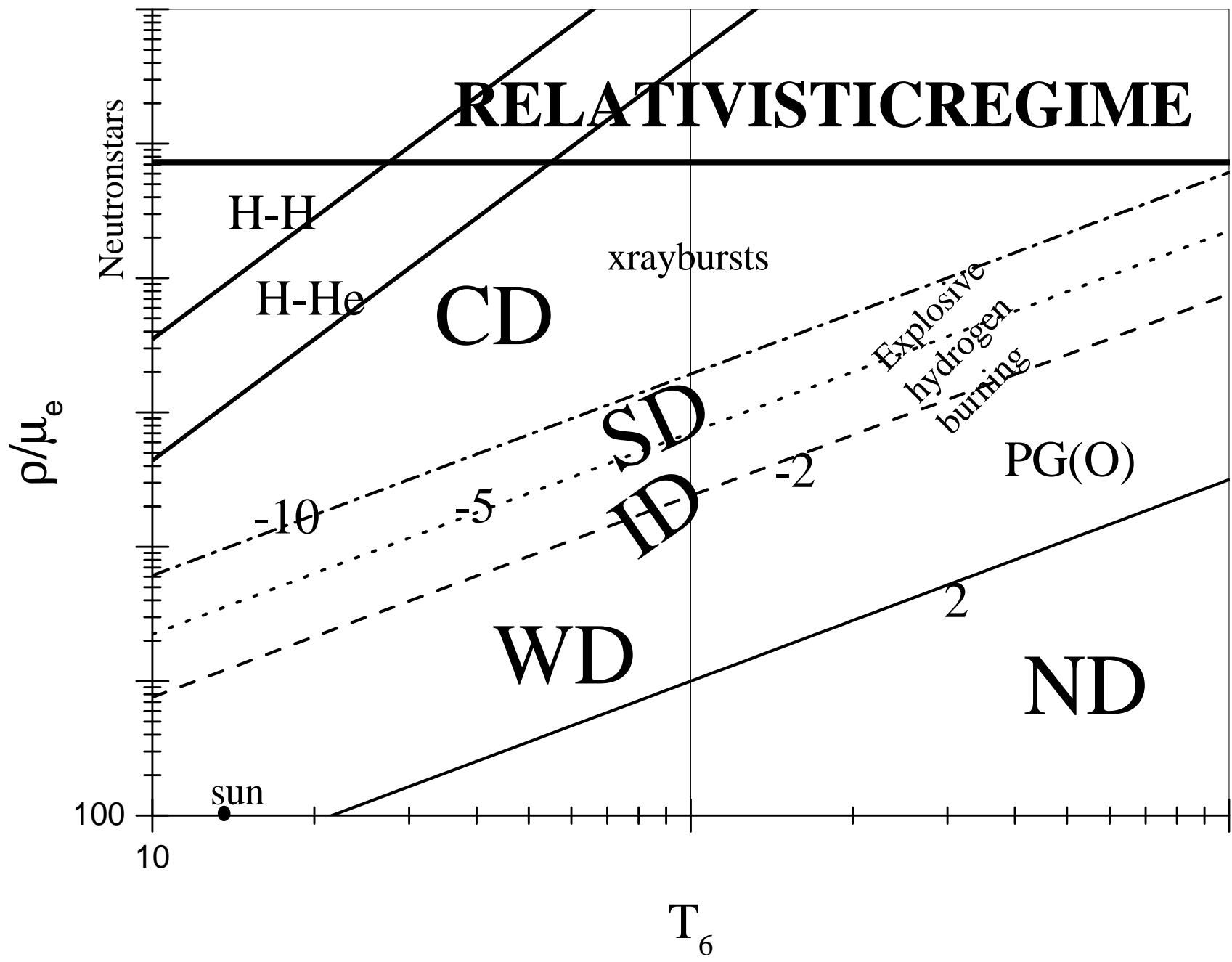
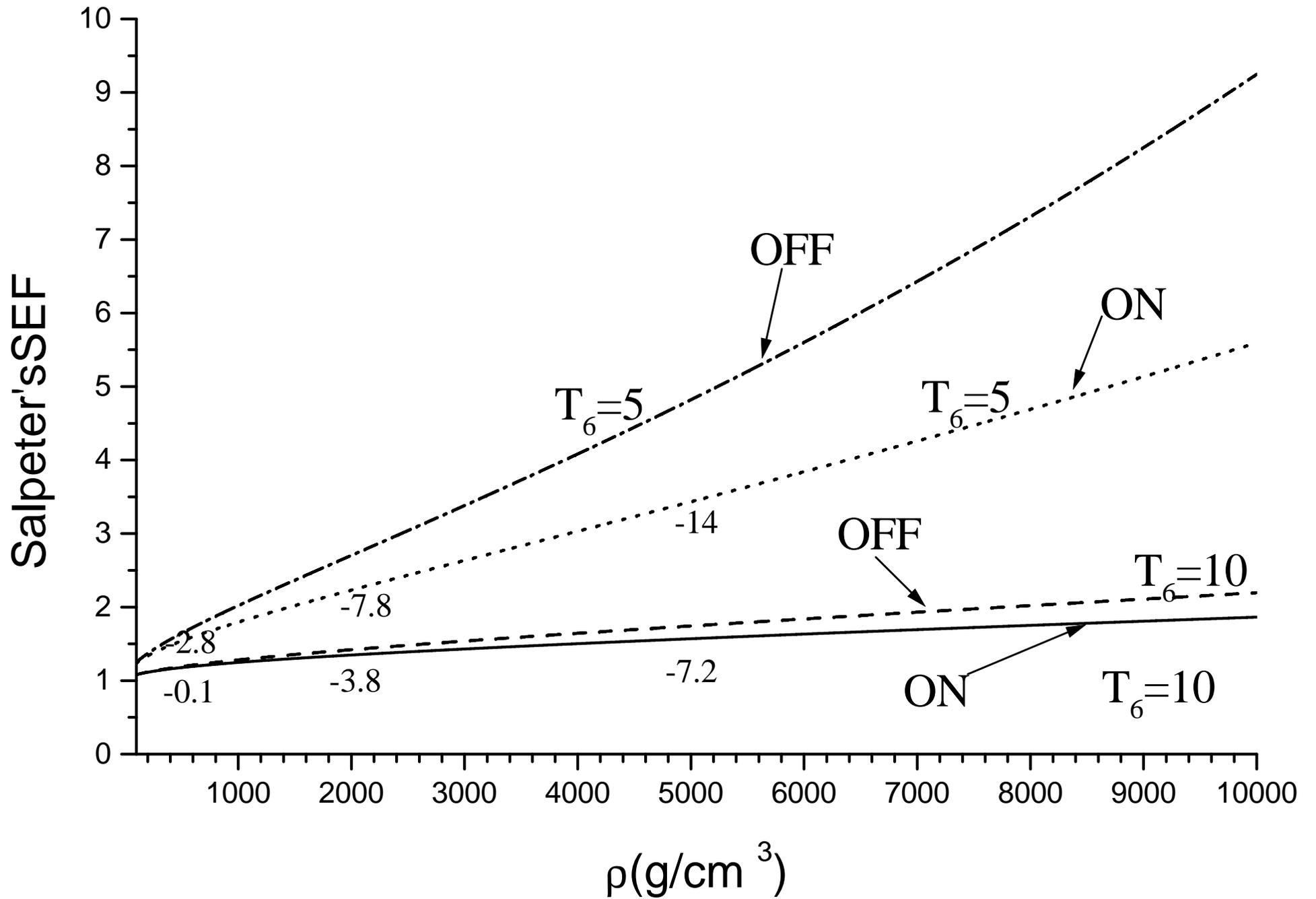
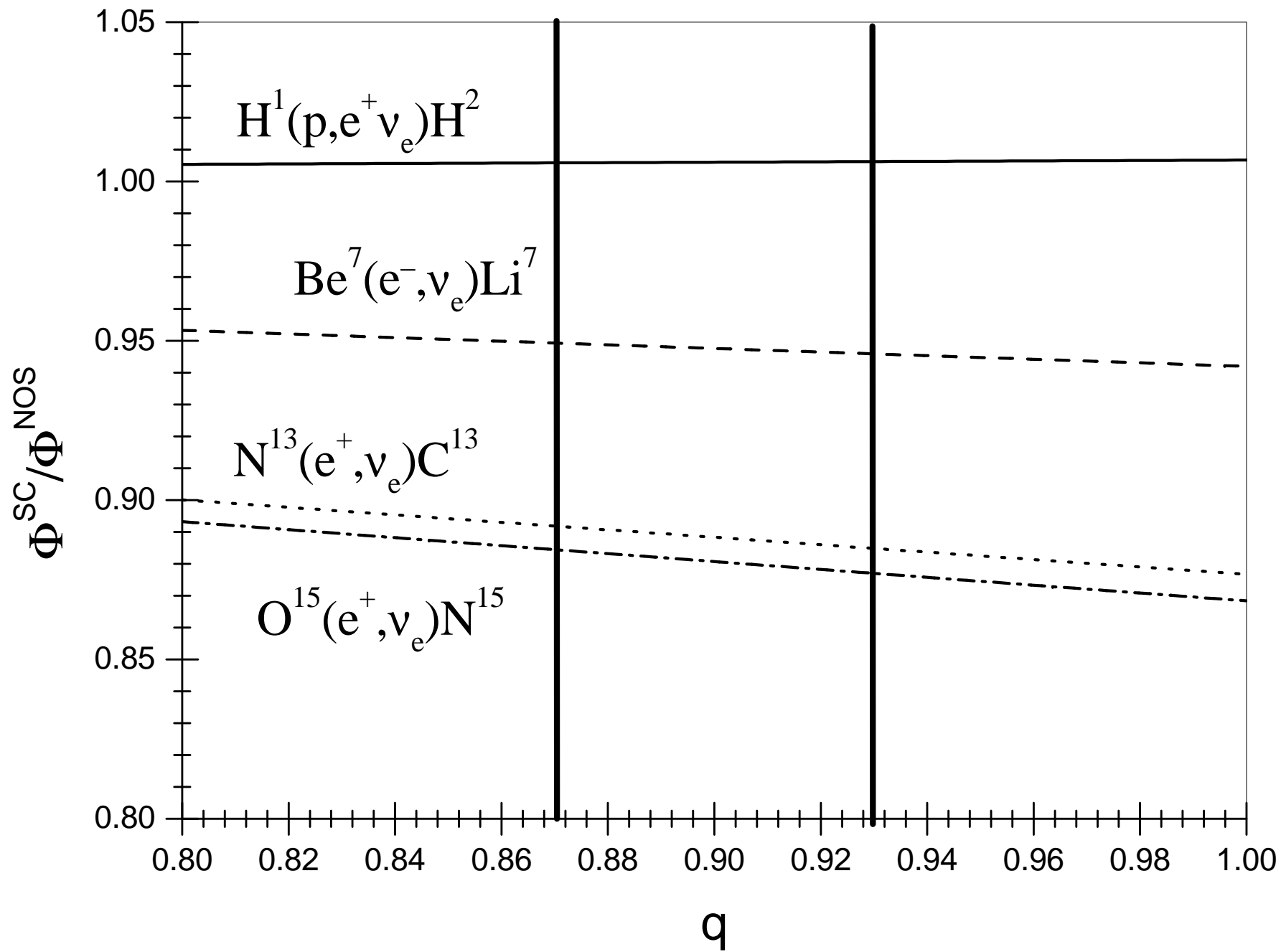
\end{document}